# Optimal Virtual Network Function Placement in Multi-Cloud Service Function Chaining Architecture

Deval Bhamare[Ψ], Mohammed Samaka[Ψ], Aiman Erbad[Ψ], Raj Jain[¥], Lav Gupta[¥], H. Anthony Chan[€]
[Ψ]Department of Computer Science and Engineering, Qatar University, Doha, Qatar
[¥]Dept. of Computer Sci. and Engineering, Washington University in St. Louis, St. Louis, MO 63130
[€]Huawei R&D USA, Plano, TX 75024
{devalb, samaka.m, aerbad}@qu.edu.qa, {jain, lavgupta}@wustl.edu, h.anthony.chan@huawei.com

**Abstract**— Service Function Chaining (SFC) is the problem of deploying various network service instances over geographically distributed data centers and providing inter-connectivity among them. The goal is to enable the network traffic to flow smoothly through the underlying network, resulting in an optimal quality of experience to the end-users. Proper chaining of network functions leads to optimal utilization of distributed resources. This has been a de-facto model in the telecom industry with network functions deployed over underlying hardware. Though this model has served the telecom industry well so far, it has been adapted mostly to suit the static behavior of network services and service demands due to the deployment of the services directly over physical resources. This results in network ossification with larger delays to the end-users, especially with the data-centric model in which the computational resources are moving closer to end users. A novel networking paradigm, Network Function Virtualization (NFV), meets the user demands dynamically and reduces operational expenses (OpEx) and capital expenditures (CapEx), by implementing network functions in the software layer known as virtual network functions (VNFs). VNFs are then interconnected to form a complete end-to-end service, also known as service function chains (SFCs). In this work, we study the problem of deploying service function chains over network function virtualized architecture. Specifically, we study virtual network function placement problem for the optimal SFC formation across geographically distributed clouds. We set up the problem of minimizing inter-cloud traffic and response time in a multi-cloud scenario as an ILP optimization problem, along with important constraints such as total deployment costs and service level agreements (SLAs). We consider link delays and computational delays in our model. The link queues are modeled as M/D/1 (single server/Poisson arrival/deterministic service times) and server queues as M/M/1 (single server/Poisson arrival/exponential service times) based on the statistical analysis. In addition, we present a novel affinity-based approach (ABA) to solve the problem for larger networks. We provide a performance comparison between the proposed heuristic and simple greedy approach (SGA) used in the state-of-the-art systems. Greedy approach has already been widely studied in the literature for the VM placement problem. Especially we compare our proposed heuristic with a greedy approach using first-fit decreasing (FFD) method. By observing the results, we conclude that the affinity-based approach for placing the service functions in the network produces better results compared against the simple greedy (FFD) approach in terms of both, total delays and total resource cost. We observe that with a little compromise (gap of less than 10% of the optimal) in the solution quality (total delays and cost), affinity-based heuristic can solve the larger problem more quickly than ILP.

## 1. INTRODUCTION

Lately there has been an exponential growth in user data traffic due to the explosion of mobile devices and the emergence of novel networking paradigms such as Internet of Things (IoT). The unprecedented increase in data traffic has resulted in excessive CapEx and OpEx for the Internet service providers (ISPs) and application service providers (ASPs) [61]. The networks built with proprietary hardware devices are complex, difficult to debug, and expensive to cater to increased demands and new emerging complex services. In addition, recently, services are moving from host-centric to data-centric model in which the computational resources are moving closer to end users. As a result, application service providers (ASPs) and ISPs are increasingly using virtualization technologies to deploy network functions over the standard high-volume infrastructure. This way of establishing network elements on the clouds is called Network Function Virtualization (NFV) [1, 2, 59].

To consolidate the gains further, the virtual infrastructure is generally obtained from cloud service providers (CSPs). In this way, users neither require knowledge, control, and ownership in the computing infrastructure nor they need to host, control or own an infrastructure in order to deploy their applications. Instead, they simply access or rent the hardware or software paying only for what they use. The possibility of paying-as-you-go along with on-demand elastic operations by cloud hosting providers is gaining popularity in the enterprise computing model. The individual functions, which were monolithic specialized hardware equipment in the past, are being replaced by a set of software-based functions called Virtualized Network Functions (VNFs) [5, 6, 55]. These functions are generally spread across multiple clouds depending on the total deployment cost, availability of the required resources and proximity to end-users [3, 4, 56].

Unlike static physical environments of the past, the task of dynamically deploying virtual network functions (VNFs) and moving them around, as the user demands change, is quite complex. VNFs are interconnected through a process called Service Function Chaining (SFC). SFC allows the formation of complex, end-to-end services by dynamically including the required network functions in the path of the traffic [3, 4, 73]. For example, a service such as "*network security*," may consist of network functions such as firewall and deep packet inspector, installed at software level. The scope of SFC is not only limited to the network services. SFC architecture is equally important for the transport services, multimedia services as well as application services. Techniques of NFV and SFC can be used by large enterprises such as banks, financial institutions, global retail stores and others to build their services in an incremental, flexible and cost-effective manner. Such enterprises are called application service providers or ASPs. For example, Netflix, Facebook or any bank having an online presence. Since more and more ASPs are embracing the multi-cloud environment, the challenge of forming and maintaining such service function chains is getting more and more complex [46].

An example of SFC is shown in Fig. 1. Let us assume a hypothetical ASP providing a service to its end-users. To fulfill the service, the user request has to go through a set of network functions, e.g., a firewall, a proxy-server, network address translator (NAT), and finally the servers implementing the business logic. These functions are deployed as VFs in the virtual infrastructure obtained from cloud service providers (CSPs) at different sites. ASPs may decide to place these functions on the available clouds based on cost, availability of required resources and proximity to users, in this case, connected via router R1, R2, R3 and R4. Hence, the user packets are routed through a service chain consisting of the following sequence of network functions: R3 (NAT), R1 (firewall), R2 (proxy server), R4 (Business Logic) and then back to the user, as shown by a red dotted line.

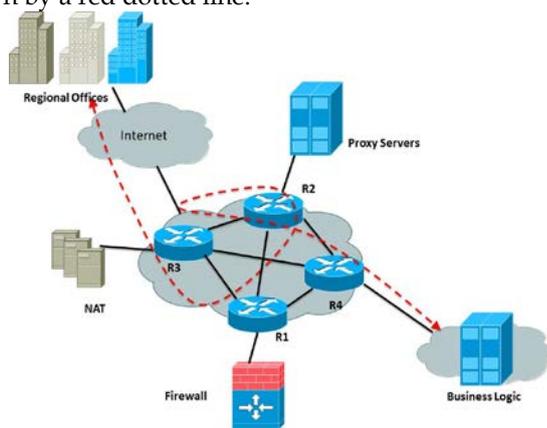

Fig. 1. Service Chain example

It is important to note that the example presented here is just an illustrative example, and a particular ASP may have several such chains depending on the functionality and business logic. The service-chains may dynamically change – e.g., grow longer and may include several branches as shown in Fig. 2. In addition, for some of the network functions, multiple instances of VNFs may need to be installed depending on the density of the user requests. Thus, the number and shapes of service function chains may vary with time and load. In addition, several chaining policies may need to be applied to meet the SLAs and QoS requirements. Hence, optimal placement of VNFs across multiple-clouds is an important problem to optimize important parameters such as network delays, network bandwidth, cost and others. Especially, with the increase in the number of the network services, network delays affect the overall performance of the composite service adversely [20]. In addition, managing resources with complex virtual function dependencies at an application level is typically ad-hoc and error-prone [49]. Considering these limitations, the Internet Engineering Task Force (IETF) SFC working group is developing an SFC architecture [8, 9, 45]. It is observed that analytical evaluation of such complex systems [12, 83] needs further studies.

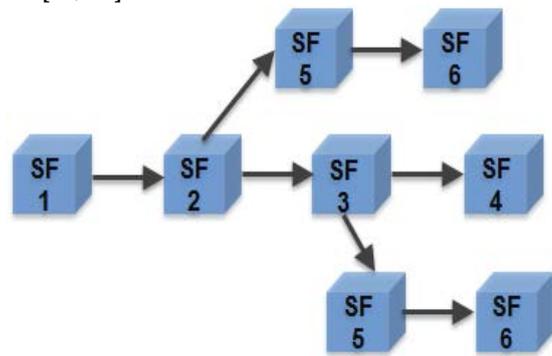

Fig. 2. Network Service, Service Functions and Service Chain.

In this work, we present an analytical model for the placement of service function chains in multi-cloud environments. Several models already exist in the literature (explained in detail in Section 2), however, most of them either focus on virtual functions (and not the chains of the functions) or consider a single cloud scenario only. In addition, the focus has been mostly on optimizing resource allocation, ignoring other important parameters such as delays to end-users, QoS and SLAs. In this work, we try to optimize end-to-end delays to end-users with optimal placement of the service function chains in a multi-cloud scenario along with important constraints such as total deployment cost and service level agreements (SLAs). We use Integer Linear Programming (ILP) method to obtain the optimal solution by setting up an objective function and applicable constraints. In addition, we model the link queues and server queues to estimate the end-to-end delays in a multi-cloud scenario accurately. Since clouds are geographically distributed and WAN links are expensive, optimizing link delays and inter-cloud traffic is an important topic for studies. With the simulation results using regression methods, we demonstrate that the M/D/1 queuing model is the most appropriate to model link queues and M/M/1 is the most appropriate for server queues [34, 35]. We only consider inter-cloud traffic since the inter-cloud links are more likely to be congested and more expensive than intra-cloud links. Intra-cloud scenarios have been studied extensively in the literature and ample amount of work is available for single-cloud environments. For such studies, the readers are requested to refer to the optimization works presented in [28, 29, 63, 68].

Further, we observe that ILP method is not scalable (beyond 100-node topology in our case) in general due to its computational complexity. Therefore, we propose a novel "Affinity-based Allocation" (ABA) heuristic approach, which solves larger problems with lesser execution time and little compromise to the solution quality. For the detailed description of the heuristic, please refer to the Section 5. We provide a performance comparison between the proposed heuristic and simple greedy approach (SGA) used in state-of-the-art systems. Common heuristics used in the state-of-art systems for the placement of VMs/SFs are "greedy with bias" [75, 77]. The bias is towards some factor such as: (1) select a service/function with the first finish or (2) select service/function with the longest finish. Similarly, the bias while selecting VMs/PMs are: (1) select most-loaded VM/PM or (2) select least-loaded VM/PM. We compare the performance of the proposed ABA approach with "Simple Greedy Allocation" (SGA) using first-fit decreasing (FFD) approach [39]. We demonstrate that with affinity-based approach, one can accommodate more stringent service level agreements (SLAs) [11, 38]. We also demonstrate that ABA produces results that are closer to optimal (gap within 10% of the optimal solution) compared against SGA, as far as total latency and total costs are considered.

The rest of the paper is organized as follows. In Section 2, we discuss the related work. An optimization model to reduce the overall latency is proposed in Section 3. We solve the model optimally using an ILP tool. We then propose ABA heuristic to solve the problem in real time scenarios for larger networks in Section 4. In Section 5, we discuss the experimental setup and present the results obtained, by comparing our novel ABA approach with the standard greedy FFD approach. Finally, we conclude the paper in Section 6.

TABLE 1
LIST OF ACRONYMS

| Acronym | Description |
|---|---|
| ABA | Affinity-based allocation |
| ASP | Application service provider |
| CAPEX | Capital expenditures |
| CDN | Content distribution network |
| CSP | Cloud service providers |
| DPI | Deep packet inspector |
| FFD | First-Fit-Decreasing |
| IaaS | Infrastructure as a service |
| ILP | Integer Linear Program |
| ISP | Internet service provider |
| NAT | Network address translator |
| NFV | Network function virtualization |
| NFVI | NFV infrastructure |
| NS | Network service |
| NSH | Network service header |
| VNF | Network virtual function |
| OF | OpenFlow |
| OPEX | Operational expenses |
| PaaS | Platform as a Service |
| SDN | Software defined networking |
| SF | Service function |
| SFC | Service function chaining |
| SFCC | Service function chaining controller |
| SFCR | Service function chaining router |
| SFF | Service function forwarder |
| SFP | Service function path |
| SGA | Simple greedy allocation |
| VM | Virtual machine |
| VNF | Virtual network function |
| VNFC | Virtual network function component |

## 2. RELATED WORK

Service Function Chaining (SFC) is an enabler for network function virtualization (NFV) networking paradigm. It provides a flexible and economical alternative to today's static environments for the Internet service providers (ISPs) and application service providers (ASPs), who use the services offered by cloud service providers (CSPs). According to [15], a service function chain is an ordered or partially ordered set of abstract service functions (SFs) and ordering constraints that must be applied to packets and/or flows selected as a result of classification. There are several working groups involved in the standardization of NFV and SFC. The standardization works are in progress at IETF, IRTF, ETSI, ITU and IEEE. ETSI formed NFV ISG (Industry Specification Group) in 2012 to define requirements and the architecture for the virtualization of network functions (e.g., Fig. 3, NFV architecture proposed by ETSI). The standards cover topics such as management and orchestration, security and trust, resilience and service quality [3-9, 40]. IETF is working on standardization of SFC architecture and its data plane elements [9, 15, 31, 45]. The Linux Foundation has launched the Open Platform for NFV Project (OPNFV) - a carrier-grade open source reference platform. OPNFV architecture supports automated, dynamic service creation and multi-domain NFV orchestration [70]. IUT-T and IEEE are also working towards the standardization of the SFC in the cloud environments [2, 16].

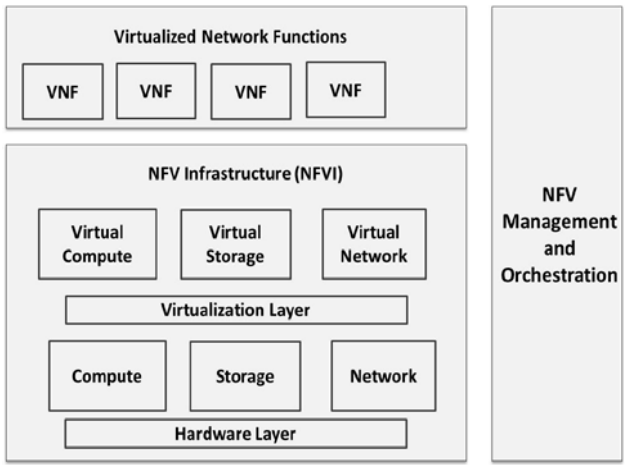

Fig. 3. NFV architecture proposed by ETSI

A system termed as "*StEERING*" has been developed using OpenFlow [48] for the practical deployment of service function chains in cloud environments [16]. A similar approach has been presented in [17, 76] as well. Mijumbi et al. provide a comprehensive survey on NFV in [53]. The authors acknowledge the fact that there is a huge scope for research in different optimization areas related to NFV, such as latency, cost, energy, network traffic and others. The authors in [55, 60, 72] provide NFV surveys from the SDN perspective. Duan and Yan present a framework for network-cloud convergence based on service-oriented network virtualization in [59]. The authors also discuss the challenges and research opportunities in network-cloud convergence. Quinn and Guichard propose an architecture based on network service headers to construct topological independent service paths needed for end-to-end service function chains [56]. Different solutions have been proposed in the literature for scheduling of network functions over virtual infrastructure, such as [63-65]. Guyton and Schwartz propose a methodology to locate the replicated services in the Internet. Similarly, authors in [71] propose a way for the description of the internet services. Authors in [83] emphasize on further research on various topics in SFCs.

Authors in [41] have proposed a model for formalizing the chaining of network functions using a context-free language. The model processes deployment requests and construct virtual network function graphs that can be mapped over the underlying network. In their opinion, NFV offers more flexibility to service function chaining by simplifying chaining and placement of VNFs. For each deployment request, the proposed heuristic chooses a VNF graph that has the minimum overall data rate requirement among all possible VNF graphs available for that request. There are automated approaches as well, such as in [16, 47], to assist the design of configurable service models, which can be applied to SFC architecture.

Wang et al. discuss optimization model for dynamic composition of the network service chains [52]. A similar approach has been presented in [27]. However, the work is limited to content distribution networks (CDNs). A distributed load management scheme using the collaborative approach in the multi-cloud environments has been proposed in [18]. The authors provide a cost-based optimization model [54] for network functions in NFV infrastructure. A similar approach has been proposed for cost optimization considering virtual machines in multi-cloud environments [58]. In [57] the authors propose an optimization model for optimal resource allocation in NFV environments. The authors in [62] provide an optimization model to reduce network traffic, however, the model needs to be modified to suit to SFC. Yoshida et al. propose a "*Multi-objective Resource Scheduling Algorithm*" (MORSA) to optimize the NFVI resources. The tool allows the NFV resource scheduler to optimize the combination of possibly conflicting objectives in complex real world situations [68]. In [69], the authors focus on the implementation of NFV over OpenFlow, especially the routing of traffic among different virtual functions.

Sonkoly et al. suggest use of virtualization techniques and propose a novel orchestration algorithm for flexible operation and optimal usage of resources [42]. The authors in [21] present cost optimization for resource subscription in multi-cloud dynamic environments. Using NFV, recent advancements in cloud computing can be leveraged and adopted in carrier environments. Flexible service definition and creation may be achieved by abstracting and formalizing the services into the concept of service chain or service graph. OpenADN is a novel approach to facilitate multi-cloud service deployment and application delivery by extending the concept of control and data plane separation proposed by the "Software-Defined Networking" (SDN) architecture [10, 14, 44].

A number of research organizations have taken up the research work in the area of SFC. However, there is a lot of work that still needs to be done to efficiently perform the placement and chaining of virtual network functions to make NFV a reality. The authors in [43] have formalized the network function placement and chaining problem and have proposed an ILP model to solve it. To make the method applicable to large deployments, they have proposed a heuristic procedure for efficiently guiding the ILP solver towards feasible, near-optimal solutions. A constrained mirror placement approach has been presented in [19] to reduce the network latency and response time in CDNs. Task scheduling algorithms have been presented for dynamic resource allocation in [22]. The goal is to reduce the required resources to perform a specific set of tasks. Dynamic resource allocation problem in cloud computing to optimize utilization of network resources and lengthy response times has been considered in [23].

Service Level Agreements (SLAs) have been part of the network industry for a long time. SLAs are getting stringent. Hence, considering SLAs in the optimization models has also become imperative as they guarantee a certain level of service performance guarantee, mostly in terms of response time, resource utilization, up-time and others. A multi-tier service model has been considered for multi-cloud environments for SLA-based optimal resource allocation in [24]. Similarly, a dynamic resource allocation problem has been considered in [25] while implementing Quality of Service (QoS). A virtual machine placement problem in the clouds while implementing SLA constraints has been considered in [26].

A significant amount of research has already been done in the context of VM placement problem, especially VM placement within a single cloud [74, 75, 77]. However, we argue that the problem needs to be revisited in the context of service function chaining. This is because, SFC architecture has some unique features, which mandates these issues to be revisited. For example, SFC is an abstracted view of the ordered service functions, which may or may not be virtual. The order in which the functions need to be visited is defined by the traffic flows dynamically. This is a unique feature of service function chains and may impose additional constraints on the already proposed solutions [83]. In addition, with scattered and geographically distributed user-bases, ASPs have been mandated to deploy the VNFs across multiple clouds. Placement of VNFs across multiple clouds is a more complex problem compared against VNF placement within a single cloud. For example, in multi-cloud scenario one has to consider link capacity constraints since WAN links are much more expensive compared against the links within a single datacenter. Hence, optimal placement of VNFs in SFCs is an important topic for the success of novel networking paradigm such as NFV and IoT.

As we observe, researchers have tried to address various optimization problems as well as have provided solutions for practical implementation of SFCs in cloud environments. Novel concepts such as network function virtualization (NFV) and software defined networking (SDN) have already been introduced to alleviate the situation [78-84]. However, concepts of NFV and SFC are relatively new and under-researched in terms of the unique challenges posed by the SFC architecture. We observe that there is a dearth of the research works which take interconnectivity between various workloads or service chains into account. Several other important optimization problems considering latency, network traffic or QoS constraints in the context of SFC for data centers or inter-cloud environments are still pending [83].

In this work, we try to formulate an analytical model for SFC architecture, by considering multiple instances of the virtual functions across multiple clouds and service chains formed due to the desired order of the flow of packets. We develop an optimization model to reduce the overall latency to the end users by trying to reduce the inter-cloud traffic. In this work, we study the architecture for the placement of service chains over multiple clouds. We set up the problem of inter-cloud traffic and response time optimization as an ILP optimization problem. The link queues are modeled as M/D/1 and server queues as M/M/1. Later on, we present affinity-based heuristic approaches to solve the problem for larger networks and provide a performance comparison with ILP. In addition, we compare our proposed heuristic with the standard greedy approach using first-fit decreasing (FFD) method.

## 3. OPTIMIZATION MODEL

In this section, we set up the problem of minimizing inter-cloud traffic and response time in a multi-cloud scenario as an ILP optimization problem. The goal of optimization model presented is to minimize the response time or latency to the clients satisfying other constraints such as the cost constraint, placement constraints (due to SLAs, explained later). We formulated the optimization model to deploy workflows on the VNFs and assign client requests to these workflows to meet the service demands. The list of variables used in the ILP is given in Table 2. Let $G = \{V, E\}$ be a graph to represent the network in consideration, where $V$ is a set of nodes representing the user-clusters in the network and E be set of the edges such that $E \in V \times V$ (Concept of user-cluster is explained in detail in Section 6, using Fig. 9). The Virtual Functions (VNFs) of the workflows will be deployed per cluster, which will be picked from the set of vertices $V$. To reduce the computational complexity of the optimization model, we compute the path between every pair of the nodes in the topology in advance, mapping paths to links. Further, we pre-calculate the delays for different traffic loads. The values for delays are stored in the matrix $T_{ij}$ as mentioned in Table 2 and selected at run-time.

The total number of sites that can be selected for deployment of VNFs, $\Gamma$, is given as an input to the optimization model. We vary this number from some minimum threshold ($\Gamma_{min}$) till maximum threshold ($\Gamma_{max}$) and observe the variation in the performance in terms of the total delay in the network.

TABLE 2
PARAMETERS FOR INTEGER LINEAR PROGRAM (ILP)

| Type | Symbol | Definition |
|---|---|---|
| Indices | i, j, k | Iterators for nodes in the topology such that $i, j, k \in |V|$ |
| | L, x | Iterator for virtual functions in the topology such that $l \in L$ |
| Input Constants | V | Set of nodes in the topology |
| | M | Total number of virtual functions a service composed of. |

|  | $\lambda_j$ | Arrival rate of packets at $j^{th}$ cloud (exponentially distributed) |
|---|---|---|
|  | $\mu_j$ | Processing rate at $j^{th}$ cloud (exponentially distributed) |
|  | $C_j$ | Computational delay at $j^{th}$ cloud. Clouds are modeled as M/M/1 model. Hence: $C_j = 1/(1- \lambda_j/\mu_j)$ |
|  | $\lambda_{ij}$ | Arrival rate of packets at link $(i,j)$ (exponentially distributed) |
|  | $\mu_{ij}$ | Processing rate of packets at link $(i,j)$ (deterministic) |
|  | $T_{ij}$ | Total delay on $(i, j)^{th}$ link/path to transmit one byte. Link queues are modeled as M/D/1 model. Hence $\Gamma_{total} = \sum_{x=1}^{L} \frac{1}{2\mu_x} \times \frac{2-(\lambda_x/\mu_x)}{1-(\lambda_x/\mu_x)}$ |
|  | $B_{ij}$ | Bandwidth of the link between $i^{th}$ and $j^{th}$ node. Value is 0 if no direct link between $i$ and $j$ |
|  | $K_j$ | Capacity vector for $j^{th}$ node (3-D vector). Value is 0 if $j^{th}$ node is a user node. |
|  | $\kappa_l$ | Capacity vector for $l^{th}$ VF (3-D vector). |
|  | $\delta_l$ | Demand vector for $l^{th}$ VF (3-D vector). |
|  | $\Delta_i$ | Demand vector for $i^{th}$ User per byte of traffic (3-D vector). Value is 0 if $i^{th}$ node is a cloud node. |
|  | $W_i$ | Traffic generated by $i^{th}$ user in number of packets. Value is 0 if $i^{th}$ node is a cloud node. Each packet size is assumed to be 500B. |
|  | $\Psi_i$ | Maximum delay per packet tolerated by user $i$ as per the SLAs |
|  | $P_i^l$ | A 2-dimensional $M \times \|V\|$ matrix. Value is 1 if $l^{th}$ function can be placed at $i^{th}$ cloud location based on the SLAs, otherwise 0. |
| Variables | $I_j^l$ | Instance matrix indicating number of instances of $l^{th}$ VF which are installed at $j^{th}$ node |
|  | $A_{ij}^l$ | Allocation matrix. Value is 1 if $i^{th}$ node (user node) is assigned to $j^{th}$ node (cloud node) for $l^{th}$ virtual function otherwise 0 |

We assume that the set of clients and clouds are disjoint sets. A cloud site $i$ has zero value for $W_i$, that is, no request flows are getting generated at clouds and only end-users can generate such flows. Similarly, a user site $i$ has zero value of $K_i$, that is, users sites do not have any processing capacities. A vector matrix $K$ represents the capacities of the sites in a vector format with $K_i = K^1_i + K^2_i + K^3_i$ being the capacity of cloud at site $i$. As mentioned earlier, we are referring to a 3-D vector to represent the capacity, that is, CPU, Storage and Network Capacity. $K_i = 0$ indicates that the site $i$ is a client site. Let $M$ be the total number of VNFs. We assume that VNFs are directly mapped to virtual machines (VMs) for their installations. For simplicity, the mapping is assumed one-to-one, hence, we may be using both the terms interchangeably. $\kappa_l$ is the vector representing capacity required for the $l^{th}$ VM. Let $\delta_l$ be the demand vector of $l^{th}$ VM and $\Delta_i$ be the demand vector for the $i^{th}$ client. For the cloud node $\Delta_i = 0$.

Let $W$ be the matrix to represent the volume of traffic originating from the client sites, that is, $W_i$ is the traffic getting generated at user node $i$. It may be noted that more than one instance of a VM may be deployed at any deployment site depending on the processing capacity of the VM and total traffic demand getting generated at the site. Let $I_j^l$ be the instance matrix representing how many instances of a VNF $l$ need to be deployed at site $j$. Let $A$ be an allocation matrix such that $A_{ij}^l = 1$ if a user at node $i$ is assigned to the cloud at node $j$. Note that $A_{ii}^l = 1$ means node $i$ has been assigned a client request. In other words, a VNF $l$ instance has been deployed on a cloud at node $i$. As mentioned in Table 2, the VF computing systems are modeled as M/M/1 queues. Using the standard formula for a M/M/1 response time, the average time spent in the system by a customer at node j is:

$$C_l = 1/(1- \lambda_l/\mu_l) \qquad (1)$$

Similarly, the links are modeled as M/D/1 queues and by the standard formula, we give the delays in the links as given in Equation (2) below. We note that $\lambda_{ij}$, that is, the link load, is a function of total flows passing through the link $(i, j)$. Hence, $\lambda_{ij}$ is computed as shown in equation (11).

$$T_{ij} = \frac{1}{2\mu_{ij}} \times \frac{2-(\lambda_{ij}/\mu_{ij})}{1-(\lambda_{ij}/\mu_{ij})} \qquad (2)$$

*Constraints*: We now discuss the constraints of the optimization model:
1. Cloud capacity: The maximum number of instances of a VNF, which may be deployed on a given cloud, is bounded by the capacity of that particular cloud and demands of the VNFs. In other words, summation of the demands of all VNFs installed in a cloud $j$ should be less than or equal to the capacity of the cloud $j$.

$$\sum_{l=1}^{M} I_j^l \times \delta_l \leq K_j \quad \forall j \in |V| \quad (3)$$

2. **VM Capacity**: The minimum number of VMs that need to be deployed on a particular cloud is bounded by fraction of the total client traffic from all the sites assigned to that particular cloud. That is, the sum of demands of clients assigned to a particular VF $k$ at a particular site $j$ should be less than or equal to the total capacity of all instances of that particular VF $k$ at site $j$.

$$\sum_{i=1}^{|V|} A_{ij}^l \times \Delta_i \times W_i \leq I_j^l \times \kappa_l, \quad \forall j \in |V|, l \in M \quad (4)$$

3. **Unity Constraint**: This constraint mandates that every client be assigned to some cloud node to get service from a VF. In addition, we assume no split of the user requests amongst the clouds for single VNF, that is, all requests from a particular user will be processed at a single cloud node only for a particular VF $l$ (single-allocation model). In other words, for a particular VF, a user should have one entry set to 1 in allocation matrix. However, we allow users to be mapped to different clouds for two different VNFs.

$$\sum_{j=1}^{|V|} A_{ij}^l = 1 \quad \forall i \in |V|, l \in M \quad (5)$$

4. **Integrity Constraint**: As mentioned earlier, we assume that the set of users and clouds are disjoint sets. Hence, we need to make sure that the user requests are forwarded to cloud nodes only (and not to the other client nodes). It is ensured with the help of following constraint:

$$A_{ij}^l \leq A_{ii}^l, \forall i, \quad j \in |V|, l \in M \quad (6)$$

5. **Cost Threshold**: The number of clouds which may be installed is an input, $\Gamma$. $\Gamma$ varies from $\Gamma_{min}$ to $\Gamma_{max}$. $\Gamma_{min}$ may start from one. However, we allow the possibility of starting with other feasible numbers. Let $f$ be the operational cost associated with a single cloud and $F$ be the total cost limit. Hence, $\Gamma_{max}$ can be calculated as $\Gamma_{max} = F/f$. At each iteration, we need to make sure that the total number of clouds hosting the VNFs in that iteration is less than or equal to $\Gamma$:

$$\sum_{i=1}^{|V|} A_{ii}^l \leq \Gamma \quad \forall l \in M \quad (7)$$

6. **Queuing Constraints**: For the queuing systems to be stable, following two constraints need to be satisfied. That is, processing rate should be greater than or equal to the arrival rate.

$$\lambda_{ij} \leq \mu_{ij} \text{ and } \lambda_j \leq \mu_j \quad (8)$$

7. **SLAs for VF Placement**: Users would normally impose a number of constraints for their services, such as, quality, operational and/or legal requirements etc. CSP and ASP sign the SLAs to meet such constraints. For example, ASP may want to deploy the firewalls at the edge locations and business logic at the core. Hence, an instance of a VF may be installed at a particular cloud location only if that location satisfies the placement constraint for that particular VF, as per the SLA. That is, a user is allocated to a cloud at node $i$ for a VF $l$ only if $l$ is allowed to be deployed at $i^{th}$ cloud as per the SLAs. The constraint may be written as follows.

$$A_{ij}^l \leq P_j^l \quad \forall i, j \in |V|, l \in M \quad (9)$$

8. **SLAs for User Response Time**: Depending on the user types (e.g., based on tariff paid or based on time sensitivity of the applications), ASPs may want to limit per packet delays for its users. This also avoids starvation of a particular user due to limited resources. However, this constraint depends on the final optimization function for total delays. Let $\Theta_i$ be the total delay for the $i^{th}$ client. The constraint can be modeled as follows.

$$\Psi_i \leq \Theta_i/W_i \quad \forall i \in |V| \quad (10)$$

9. **Multi-Cloud Link Delays**: This constraint models the link delays as a function of total traffic passing through the link. This is important as the link flows are not static and vary as more and more clients are added to the network. Various models have been proposed to model the total traffic and link parameters; however, we consider the stochastic model for the link delays.

$$\lambda_{ij} = \sum_{k \in |V|} \sum_{l \in M} W_k \times (A_{ki}^l \times A_{kj}^{l+1})$$
$$\forall i, j \in /V| \quad (11)$$

***Optimization Function***: We seek to minimize the total response time to the end-users in the network. Delays are divided into two categories: transmission delays associated with links and computational delays associated with the clouds. Term $A_{ki}^l \times A_{kj}^{l+1}$ confirms that $k^{th}$ user is assigned to a cloud at node $i$ for $l^{th}$ VF and to a cloud at node $j$ for $(l+1)^{th}$

VF. If so, then we multiply the term with the transmission delay between nodes $i$ and $j$ as well as computational delay at node $j$ ($T_{ij}$ and $C_j$, respectively). For a connection between a user and the very first VF in the service chain, we have a separate case, which is the first term in the optimization function. For the sake of simplicity, we assume that the VFs are visited in the numerical order. There may exist different service flows following different chains, however, the numbers for the VFs are in numerical order. For example, different chains consisting of different VFs may exist, such as (1,2,3,4), (1,2,3,5), (6,7,9,10), (15, 17, 20) and others as shown in Fig. 4.

We solve the ILP formulation using *Integer Linear Program* (ILP) tool. We formulate our optimization function as follows.

Minimize:
$$\sum_{k \varepsilon |V|} \sum_{j \varepsilon |V|} A^1_{kj} (T_{kj} + C_j) + \sum_{l \varepsilon M} \sum_{k \varepsilon |V|} \sum_{i \varepsilon |V|} \sum_{j \varepsilon |V|} A^l_{ki} \times A^{l+1}_{kj} (T_{ij} + C_j) \quad (12)$$

*Linearization of ILP*: We formulate an optimization function as shown in (12) above. However, we notice a non-linearity in the equation due to multiplication of $A^l_{ki}$ and $A^{l+1}_{kj}$. To remove the non-linearity, we introduce another variable $\Phi^l_{ij}$ such that:

$\Phi^l_{ij} = 1$ iff $A^l_{kj}$ and $A^{l+1}_{kj}$ = 1, otherwise 0  (13)

satisfying the constraints below:
$\Phi^l_{ij} \leq A^l_{ki}$ and $\Phi^l_{ij} \leq A^{l+1}_{ki}$  (14)
$\Phi^l_{ij} \geq A^l_{ki} + A^{l+1}_{ki} - 1$  (15)

The optimization function may be re-written as:

$$\sum_{k \varepsilon |V|} \sum_{j \varepsilon |V|} A^1_{kj} (T_{kj} + C_j) + \sum_{l \varepsilon M} \sum_{i \varepsilon |V|} \sum_{j \varepsilon |V|} \Phi^l_{ij} (T_{ij} + C_j) \quad (16)$$

The results obtained after solving the ILP are presented in Section 6. The computational complexity of the optimization model is very high. We note that $A$ is a 3-dimensional matrix. Due to the term $A^l_{ki} \times A^{l+1}_{kj}$, the total complexity of the ILP is $O(V^4M^2)$, where $V$ is the total number of users' nodes and M is the total number of virtual functions. As $M << V$, the complexity may be written as $O(V^4)$, which is still very high. Due to this high computational complexity, application of this optimization may be restricted to small data sets. Hence, we propose heuristic approaches in the next section to solve real time problem for larger number of users.

## 4. HEURISTICS

The problem under consideration is a two-fold problem. The first part consists of placing the VNFs in the clouds while the second part consists of allocating the user flows to the already placed VNFs. In this work, we present a novel heuristic approach to solve the aforementioned problem. The proposed approach involves "*affinity-based*" allocation (ABA). It takes into account the traffic propensity among the VNFs while placing the VNFs in clouds. We compare our proposed heuristic with a standard "*greedy*" method [19] to place the VNFs on the clouds. Especially we consider a simple greedy approach (SGA) using FFD (first-fit decreasing) method, which is prominent in the literature. In this method the VNFs are organized in a decreasing order of resource requirement and placed on physical resources arbitrarily 'opening' a new physical server if the next VF requires more resources than available in any of the available servers [36-39]. We observe that the greedy approach produces results comparatively quickly than that of affinity-based approach. However, the solution quality is much better and closer to the optimal with the affinity-based approach (margin of less than 10% of the optimal solution, as explained in Section 6).

In the greedy approach, we first determine the instances of all VNFs, which will be needed to satisfy all user demands. We have considered placement constraints imposed by SLAs (such as, some of the VNFs has to be placed at core sites and others at the edge locations). These are discussed in detail later in this section. The greedy approach continues iterating sequentially through all the instances of all the VNFs to place them on the appropriate clouds, satisfying the capacity constraints. VNFs are placed on the appropriate cloud using the greedy approach, that is, the heuristic tries to fit as many as possible VNFs on a single cloud before it moves to the next one. Table 3 describes the steps in reading all input parameters and performing the pre-processing step. This step is common to both heuristics presented in this work.

TABLE 3
HEURISTIC STEP I

| *Input Parameters and Initial Construction* |
|---|
| 1. Read **λ** and **µ** for each link for average load conditions as Input Parameters |
| 2. Read **Service Graph** and **% of flows** between VNFs   //traffic based affinity matrix |
| 3. Construct Matrix **T (V × V)**   //Placement Affinity matrix |
| 4. Read Matrix **P(V × 2)**   //N = number of user users |
| 5. Read **W(N × 1)** as User Weight matrix   // M = number of clouds |
| 6. Read **C(M× 1)** as Cloud Capacity matrix   //Number of instances of each VF installed on each cloud |
| 7. Construct Instance Matrix **I( V × M)** |

| | |
|---|---|
| | //Delay each User can tolerate |
| 8. | Read Delay Matrix **D(N)** for each user |
| | //Cost threshold |
| 9. | Read Cost Threshold Matrix **S(N)** for each user |

To incorporate the placement constraints, we divide the clouds into edge-clouds and core-clouds. Edge-clouds are located at the periphery of the topology and closer to the end-users or user-clusters. In this work, we consider five SFCs comprising of twenty-five VFs in total. The SFC shapes and graphs are given in the Fig. 4 and Fig. 5. We note that the shapes of the SFCs also indicate their execution order. For example, in SFC 1, VNF $f_2$ has to be executed after $f_1$. This may be due to the business logic dependence or some mandatory network traffic flow demands. For example, VNF handling the web-service logic has to be placed before VNF handling databases or firewall must be placed before the business logic, etc. For a detailed understanding of the service flow, provided by a hypothetical ASP, we investigate fifth SFC in more depth as shown in Fig. 5. This particular SFC comprises of five VFs. These functions may be business logic, DPI, Firewall, NAT and Database (numbered as 1 through 5, respectively).

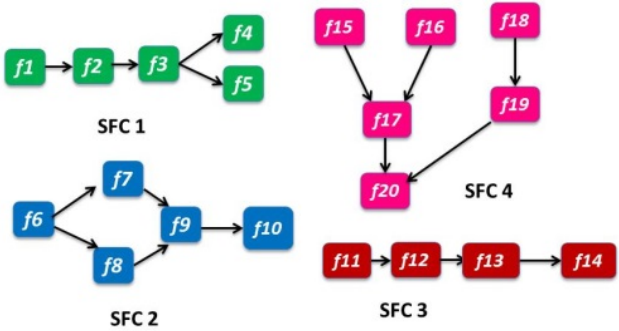

Fig. 4. Service Function Chains.

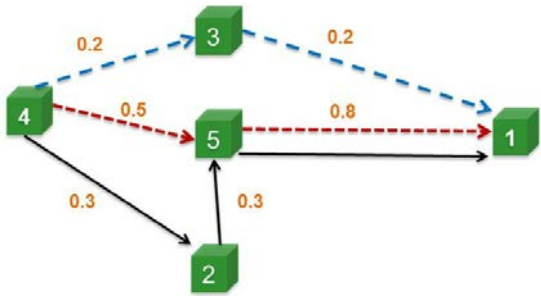

Fig. 5. A Service Flow Graph for SFC 5.

In this case, the VNF for business logic has to be at the core locations (not exposed to the users). On the contrary, NAT has to be on the edge sites, closer to the end users and not at the core sites. Other functionalities (VNFs) may be deployed at any location as per the resource availability [13] or proximity to the end-users. It may be noted that these requirements may change as per the rules and policies of the ASPs and SLAs. The corresponding placement constraint matrix for the above example is given in Table 4. Value 1 in the site columns indicates that the particular VNF has to be deployed at that particular location, while entry of -1 indicates that the instance of the particular VNF cannot be installed at that site. 0 in the table indicates "*don't care*" condition. The traffic-affinity matrix based on the service graph of Fig. 5 is given in Table 5. It represents the traffic flow among the five VNFs of SFC 5 as a fraction of the total traffic. The traffic affinity is taken into consideration while placing the VNFs in the "*affinity-based allocation*" (ABA) approach. It is to be noted that the placement constraints and traffic affinity constraints are applicable to all SFCs under consideration.

TABLE 4
VNF PLACEMENT CONSTRAINTS FOR SFC 5

| Number of VNFs | Core Site | Edge Site |
|---|---|---|
| 1 | 1 | -1 |
| 2 | 0 | 0 |
| 3 | 0 | 0 |
| 4 | -1 | 1 |
| 5 | 0 | 0 |

TABLE 5
FRACTION OF DATA FLOWS BETWEEN THE VNFS

| VNFs | 1 | 2 | 3 | 4 | 5 |
|---|---|---|---|---|---|
| 1 | 0 | 0 | 0 | 0 | 0 |
| 2 | 0 | 0 | 0 | 0 | 0.3 |
| 3 | 0.2 | 0 | 0 | 0 | 0 |
| 4 | 0 | 0.2 | 0.3 | 0 | 0.5 |
| 5 | 0.8 | 0 | 0 | 0 | 0 |

Table 6 shows the steps for the greedy heuristic. As mentioned earlier, greedy approach iterates through all the instances of all VNFs. Later on, the heuristic iterates through all users to allocate them to the appropriate cloud.

TABLE 6
GREEDY HEURISTIC STEP II

| |
|---|
| ***VF Location and Users' Allocation***: |
| 1. //Traverse sequentially through the list |
| 2. Foreach (VF ***v*** in ***V***) |
| 3.     Foreach (cloud ***m*** in ***C***) |
| 4.     { |
| 5.         //Placement constraints are satisfied and cloud has //capacity |
| 6.             if ( P(v, c) != -1 and P(v′, c) != -1 and C(m) >= D(v + v) |
| 7.                 Install instances of ***v*** at ***m*** |
| 8.             Repeat until |
| 9.                 All instances of VF ***v*** are installed on m or capacity of m is exhausted |
| 10.     } |

|  |  |
|---|---|
| 11. | Foreach(User **u** in **U**) |
| 12. | Foreach (cloud **m** in **C**) |
| 13. | { |
| 14. | //Sufficient VNF instances are installed on //the cloud |
| 15. | if (I(u, v) >= u¹ and D(u, c) <= Γ) |
| 16. | Allocate user **u** to cloud **m** |
| 17. | } |

In this work, we have considered three user-classes, namely (1) Gold, (2) Silver and (3) Bronze; depending on the tariffs paid by the users. Higher the tariff better is the service offered to the user. Gold users pay highest tariffs and should suffer minimum delays among the three classes of the users. On the contrary, bronze users pay the least tariffs and may be subjected to a longer delay compared to the other two classes (Table 7).

TABLE 7
USER TYPES

| User Type | Gold | Silver | Bronze |
|---|---|---|---|
| Tariff | Highest | Medium | Lowest |
| Delay Tolerance | Lowest | Medium | Highest |

While allocating a particular user flow to the cloud, we first ensure that the cloud has sufficient instances of the required VNFs installed. In addition, we make sure that the predicted delays for the given class of the user are below the user's tolerance level (as per the user classes, Table 7). Heuristic is provided with the statistical data from the ASPs for the average packet arrival rates ($\lambda$) and link processing rates ($\mu$) for each link in the network. As mentioned earlier, the links have been modeled as *M/D/1* models (Section 4). Hence, the total processing time for a packet on a particular link can be given as shown in Equation 2. Since the links are *M/D/1* model, the link delays can be added to get the total delays [66, 67].

Once a user is allocated to the cloud to get the desired service, heuristic predicts the network delays for that user. This is explained with a simple example. As shown in Fig. 6, if a user is allocated to the cloud 3, the links on that particular path are identified (generally we choose k-shortest paths). In this case, the links under consideration are *L1* and *L2*. The values for $\lambda$ and $\mu$ are read from the input and the total delays are calculated as shown in Equation 17. Note that $x$ is the iterator used to iterate through the links present in the path under consideration. If the total delay is greater than the tolerance limit of the user, next shortest path is chosen. This process is repeated till all k-shortest paths are exhausted. If no such path is found, user is allocated to the next feasible cloud. In addition, we consider the computational delays at the clouds while calculating total delays.

$$\Gamma_{total} = \sum_{x=1}^{L} \frac{1}{2\mu_x} \times \frac{2-(\lambda_x/\mu_x)}{1-(\lambda_x/\mu_x)} \quad (17)$$

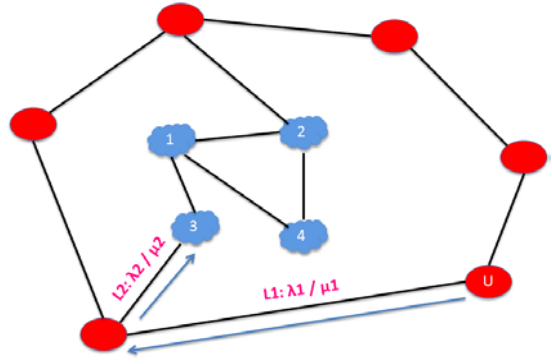

Fig. 6. Predicting User Delays.

Affinity-based allocation approach (ABA) considers the affinity between the VNFs while placing the VNFs on the clouds. The logic is instances of the VNFs should be installed closer to each other if the affinity among the VNFs is more, ideally on the same cloud. In this case, we have considered traffic-based affinity between the VNFs (Table 5). The higher is the traffic between the VNFs, the greater is the affinity between them. The intuitive logic is that, if we place the instances of the VNFs with more traffic-affinity on the same cloud, the total inter-cloud traffic and hence end-to-end delays will be less. The service graph and percentage of traffic flows, given in Fig. 5, represent a sample service provided by an ASP.

The service consists of five different functions as mentioned earlier in the section. These functions may be Firewall, NAT, DPI, Database and business logic. These functions are implemented as VNFs. A CSP may deploy as many instances of these VNFs as needed over its IaaS infrastructure. Each user request has to travel through the NAT virtual function first (indicated by block 4). Let there be three different traffic flows. First flow is through virtual functions or blocks 4→2→5→1 (black solid lines). Let us assume that this traffic requires segregation based on the type of payload and is, therefore, required to go through the DPI function (block 2) before going through the database (block 5) and finally to business logic (block 1). Let us assume that ASP statistics indicate that generally 30% of the user requests need to go through this specific path. Next 20% traffic follows on path 4→3→1 (blue dashed lines), with block 3 as a firewall function. Remaining 50% traffic goes directly through database to business logic following the path 4→5→1 (red dotted lines). Please note that actual ASP service flows may be more complex, however, we have considered this "five VNFs" SFC case for better understanding.

The first part of the affinity-based heuristic is similar to the greedy one as shown in Table 3 earlier. The only difference is that we additionally calculate the affinity matrix for the VNFs. The matrix indicates the fraction of the total traffic that will flow among the corresponding VNFs. For the VNFs' and users' allocation (that is the second step of the heuristic), the steps involved in ABA approach are given in Table 8. The execution complexity of the affinity-based approach may be expressed as $O(N \times M \times V)$, where $N$ is the total number of users, $M$ is the number of clouds and $V$ is the number of virtual functions which form the service of a hypothetical ASP. If we consider that $M \& V << N$, we observe that complexity turns out to be linear, that is, $O(N)$. Experimental setup and the results obtained for the ILP and proposed heuristics are explained in the next section.

TABLE 8
AFFINITY-BASED HEURISTIC STEP II

| VF Location and Users' Allocation: |
|---|
| 18. While (not all VNFs *v* in **V** are considered) |
| 19. { |
| 20.   Let *v* = next VF with highest traffic-affinity value in **T** |
|       //We give priority to VNFs which have traffic affinity |
| 21.   Let *v'* be the VF s.t. **T**(*v*, *v'*) is next highest |
| 22.   Foreach (VF *v'* in **V**) |
| 23.    Foreach (cloud *m* in **C**) |
| 24.    { |
|        //Check if placement constraints allow the placement |
|        //and cloud has the capacity |
| 25.    If ( P(v, c) == 1 \|\| P(v', c) == 1) && C(m) >= D(v + v)) |
| 26.       Install instances of *v* and *v'* at *m* |
| 27.    If (P(v, c) != -1 && P(v',c)!=-1) && C(m) >= D(v+ v)) |
| 28.       Install instances of *v* and *v'* at *m* |
| 29.    Repeat until all instances of *v* and *v'* are installed on *m* or capacity of *m* is exhausted |
| 30.   } |
| 31. } |
| 32. Foreach (User *n* in **N**) |
| 33.   Foreach (cloud *m* in **M**) |
| 34.    Foreach (VF *v* in **V**) |
|        //if delay at cloud m for user n is acceptable |
| 35.     If ( $\Gamma(n \times m) <= D(N)$ ) |
| 36.      If( $I(v \times m) > 0$ ) |
| 37.      { |
|          //user n is allocated to cloud m for VF v |
| 38.       $X_{nm}^{v} = 1$ |
|          //reduce available capacity of m |
| 39.       Update C (m) |
| 40.      } |

## 5. EXPERIMENTAL SETUP AND RESULTS

In this section, we analyze the performance of the proposed affinity-based (ABA) heuristics against simple greedy approach using first-fit decreasing (FFD) method. We compare their results with the results of the ILP based solution. In addition, we compare the results of the greedy (FFD) approach with the affinity-based (ABA) approach.

Due to the computational complexities, ILP seems to be suited to problems with smaller instances. However, we demonstrate that with the proposed ABA heuristic, larger sets of problems can be solved with a little compromise in the solution quality. With little loss of quality, we can obtain a greater applicability of the ILP, especially for larger networks and quicker solutions. For example, on a quad-core 2.7 GHz processor, ILP took around 500 seconds and 4 GB RAM (random access memory) for 10-cluster topology; while for 100 clusters, time taken was around 5000 seconds with 30 GB RAM. On the contrary, both heuristics take less than 500 seconds and less than 4 GB RAM for up to 1000 clusters. We have used the following resource configurations (from Amazon EC2 [50]) to simplify configurations so that resource requirements can easily be mapped to the nearest available configuration. Depending on the VNFs, a particular VM is chosen from Table 9, so that resource requirements can easily be mapped to the nearest available configuration. A particular VM is chosen from Table 9 such that the requirements are the closest match. We may combine two or more VNFs and deploy them on a single VM as well, provided a VM of the required capacity is available. The availability of the VMs depends on the total cloud capacity.

TABLE 9
RESOURCE CONFIGURATION FROM EC2

| Configuration | Memory (GB) | Compute Unit | Disk (GB) | Platform (bit) | Cost ($/h) |
|---|---|---|---|---|---|
| m1.small | 1.7 | 1 | 160 | 32 | 0.1 |
| m1.large | 7.5 | 4 | 850 | 64 | 0.4 |
| m1.xlarge | 15 | 8 | 1690 | 64 | 0.8 |
| c1.medium | 1.7 | 5 | 350 | 32 | 0.2 |
| c1.xlarge | 7 | 20 | 1690 | 64 | 0.8 |

We have used GUESS software [51] to generate random graphs for testing. GUESS is an open-source data analysis and visualization tool for graphs. A sample graph with 200 user clusters (blue squares) and ten (5% of the total nodes in the network) clouds (red circles) is shown in the Fig. 7 [33]. A user cluster here refers to an ISP network with a 3-tier hierarchy of routers: access, aggregation and core routers. A detailed structure of a sample user cluster is shown in Fig. 8. A single ISP network consists of several users, which are connected to the access routers. On an average, we have considered 1000 users per cluster. For each cluster, traffic is aggregated at the aggregation router and then passed to the next hop. Access routers are eventually connected to the aggregation router. The aggregation router either routes the traffic to the cloud for processing or to the core router, through which it eventually reaches the cloud for processing.

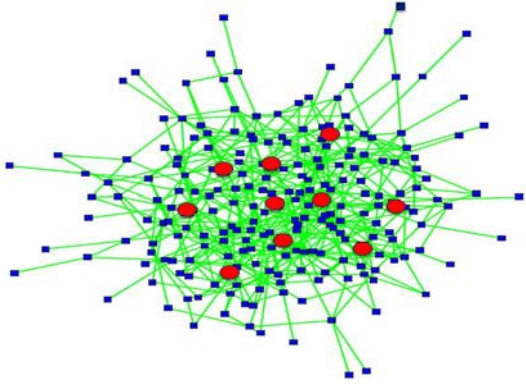

Fig. 7. A sample 200 node topology generated using GUESS.

In the rest of the article, we have used the term "user cluster" to represent aggregator node of ISP network as explained above. We assume that the aggregation routers are service-chain aware routers. In other words, these routers have SFFs implemented to differentiate among the user flows as per the class and find out the exact path in the chain the user flows need to follow [3, 6]. These SFFs can easily be implemented in the application layer with SDN [32].

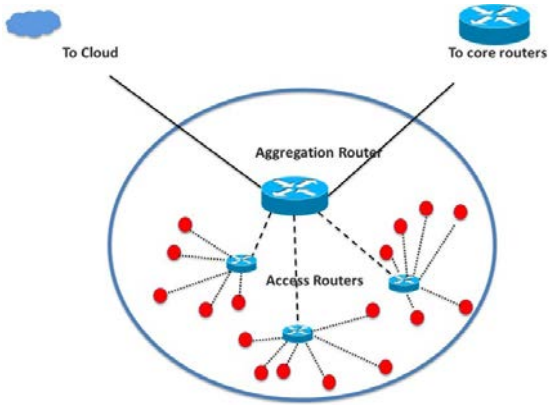

Fig. 8. A closer look at the user-cluster.

We measure the total time required to satisfy all the user demands. A user demand is successfully met if the user request traverses through the predefined set of virtual functions in a given order and the response generated at the last VNF reaches back to the user as an acknowledgment within acceptable time limits. It is to be noted that, while optimizing the total delays, we make sure that the SLA constraints for cost, delays and affinity for every user are also satisfied. We obtain results for different topology sizes by varying the user-cluster sizes and traffic loads. For simplicity, we assume each packet has a size of 500 bytes. The packet generation rate is varied to simulate different traffic rates.

The link capacities are assumed to be 100 Kbps, {1, 10, 100, 255} Mbps or 1 Gbps, chosen randomly. For the experimental setup, we have considered a closed-loop system. One user request is assumed to be a set of 50 data packets. For this one set of packets or one user request, a single reply is sent back by the cloud to the user as an acknowledgment for the request completion. The next request is sent only after reply to the previous one is received. Every user sends a predefined amount of data, selected randomly. Depending on the desired rate of transmission, the user sends data at a specific rate. For example, if $k^{th}$ user has 10 GB of data to send, then that particular user will generate $2 \times 10^7$ packets in total since the packet size is assumed to be 500 bytes. In addition, we assume that the number of clouds in the network is 5% to 20% of the user-clusters, depending upon total user-base size. Below we present the results obtained with the ILP as well as the proposed heuristics, using the experimental setup. We measure the total time required for all users in the system to get their requests satisfied. We plot the graphs for total delays against the total number of users in the system as well as the total traffic load on the system. The results and observations are discussed next.

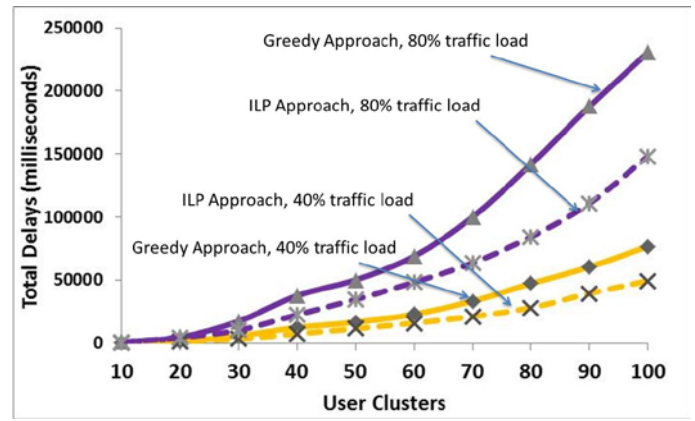

Fig. 9. ILP vs. FFD Greedy (varying cluster size).

Graphs presented in Fig. 9 display the total delays obtained using ILP (dashed lines) and the greedy approach (solid lines) with constant traffic loads and varying user-cluster numbers along X-axis. Due to the computational complexity of the ILP, we have considered topologies with the number of clusters varying from 10 to 100 only. We obtain results for 40% and 80% traffic loads in the network. We observe the expected growth in the total response time as the number of clusters increases. In addition, total response time at 80% traffic load is higher than that of at 40%. This is due to the fact that the queuing delays increase as the traffic load increases. However, we observe that the quality of the solution generated by a standard FFD greedy approach is degraded. The same behavior is observed in Fig. 10 where we have plotted the delays against varying traffic loads while the number of clusters is kept constant. The gap between the optimal solution and the greedy approach keeps on increasing as the problem size goes on increasing. We observe a gap of almost 30% to 40% at 80% traffic load and cluster size as 90. Detailed results for comparison between ILP and Greedy approaches are presented in Table 10. In Fig. 11, we have plotted column charts for the comparison between ILP and FFD Greedy heuristic. The readings are taken at 60% and 70% traffic loads. We clearly observe the larger gap between the greedy heuristic and optimal solution in Fig. 11.

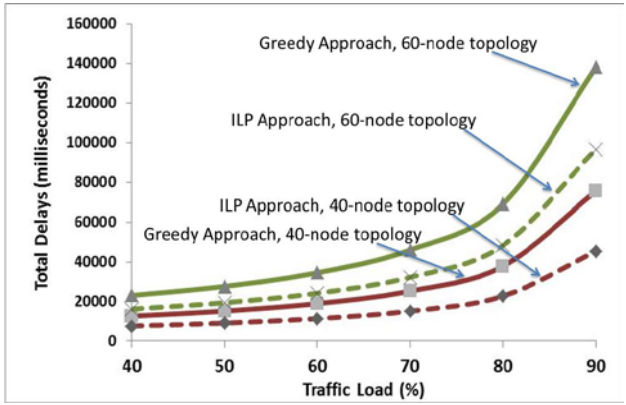

Fig. 10. ILP vs. FFD Greedy (varying traffic load).

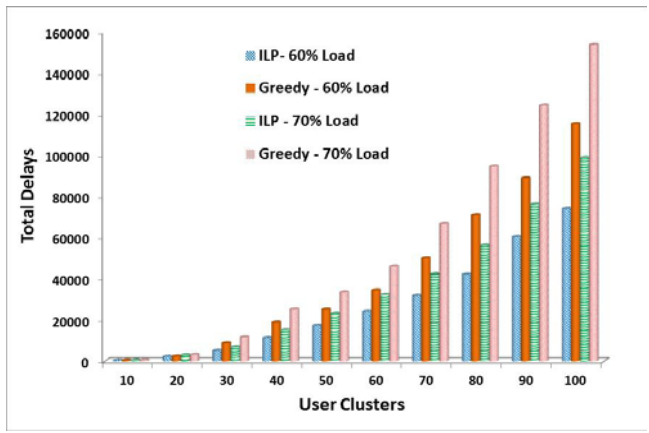

Fig. 11. ILP vs. FFD Greedy column-chart (varying user cluster size).

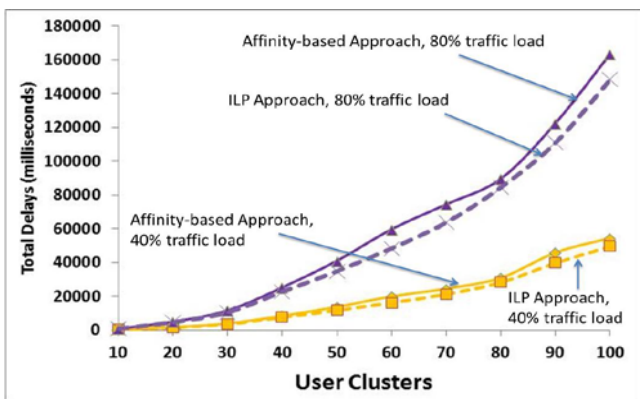

Fig. 12. ILP vs. Affinity-based heuristic (varying cluster size).

Graphs presented in Fig. 12 and Fig. 13 display the total delays obtained using ILP (dashed lines) and the affinity-based approach (ABA) (solid lines) with other parameters are as explained above for Fig. 9 and 10, respectively. We observe close to optimal performance with much reduced response time using the ABA scheme. Since the heuristic tries to place the virtual functions with more traffic flows in the same cloud, inter-cloud traffic is much reduced, which eventually reduces the total response time. The gap between the results of the optimal solution and ABA scheme has been observed to be less than 10%.

TABLE 10
ILP VS. FFD GREEDY

| Number of User Clusters | Traffic Load (%) | | | | | |
|---|---|---|---|---|---|---|
| | 40(%) | 50(%) | 60(%) | 70(%) | 80(%) | 90(%) |
| | ILP Delays (milliseconds) | | | | | |
| 10 | 260 | 312 | 390 | 520 | 780 | 1560 |
| 20 | 1520 | 1824 | 2280 | 3040 | 4560 | 9120 |
| 30 | 3456.67 | 4148 | 5185 | 6913.33 | 10370 | 20740 |
| 40 | 7553.33 | 9064 | 11330 | 15106.7 | 22660 | 45320 |
| 50 | 11593.3 | 13912 | 17390 | 23186.7 | 34780 | 69560 |
| 60 | 16110 | 19332 | 24165 | 32220 | 48330 | 96660 |
| 70 | 21213.3 | 25456 | 31820 | 42426.7 | 63640 | 127280 |
| 80 | 28146.7 | 33776 | 42220 | 56293.3 | 84440 | 168880 |
| 90 | 39626.7 | 41352 | 60440 | 76253.3 | 110880 | 231760 |
| 100 | 49423.3 | 59308 | 74135 | 98846.7 | 148270 | 296540 |
| | FIFO Greedy Delays (milliseconds) | | | | | |
| 10 | 316.667 | 380 | 475 | 633.333 | 950 | 1900 |
| 20 | 1600 | 1920 | 2400 | 3200 | 4800 | 9600 |
| 30 | 5850 | 7020 | 8775 | 11700 | 17550 | 35100 |
| 40 | 12600 | 15120 | 18900 | 25200 | 37800 | 75600 |
| 50 | 16750 | 20100 | 25125 | 33500 | 50250 | 100500 |
| 60 | 23000 | 27600 | 34500 | 46000 | 69000 | 138000 |
| 70 | 33366.7 | 40040 | 50050 | 66733.3 | 100100 | 200200 |
| 80 | 47333.3 | 56800 | 71000 | 94666.7 | 142000 | 284000 |
| 90 | 60600 | 74120 | 88900 | 124200 | 187800 | 375600 |
| 100 | 76833.3 | 92200 | 115250 | 153667 | 230500 | 461000 |

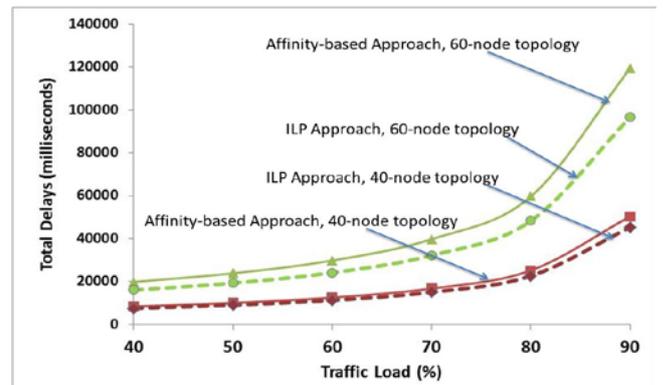

Fig. 13. ILP vs. Affinity-based heuristic (varying traffic loads).

For better insights, in Fig. 14 we plot the column-chart for the comparison between ILP and Affinity-based heuristic. We observe a reduced gap between optimal solution and heuristic solution using the proposed Affinity-based approach (ABA). Table 11 represents the detailed results of comparison between ILP and Affinity-based approaches.

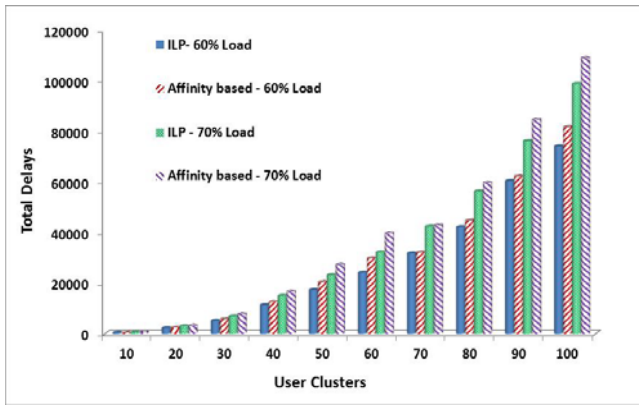

Fig. 14. ILP vs. Affinity-based heuristic column-chart (varying cluster size).

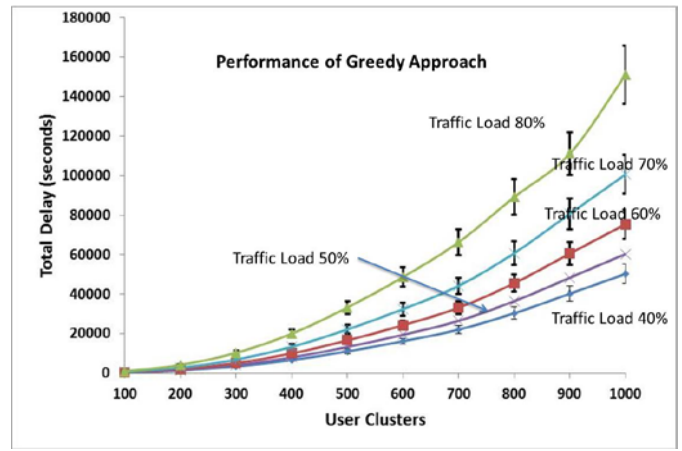

Fig. 15. Performance of FFD Greedy.

Fig. 15 displays the performance of the FFD Greedy approach as the user cluster size varies from 100 to 1000. The readings are taken at traffic loads of 40% to 80%. Similarly, we present the graphs to demonstrate the performance of the ABA scheme in Fig. 16 with similar setup.

TABLE 11
ILP VS. AFFINITY-BASED

| Number of User Clusters | Traffic Load (%) | | | | | |
|---|---|---|---|---|---|---|
| | 40(%) | 50(%) | 60(%) | 70(%) | 80(%) | 90(%) |
| | ILP Delays (milliseconds) | | | | | |
| 10 | 260 | 312 | 390 | 520 | 780 | 1560 |
| 20 | 1520 | 1824 | 2280 | 3040 | 4560 | 9120 |
| 30 | 3456.67 | 4148 | 5185 | 6913.33 | 10370 | 20740 |
| 40 | 7553.33 | 9064 | 11330 | 15106.7 | 22660 | 45320 |
| 50 | 11593.3 | 13912 | 17390 | 23186.7 | 34780 | 69560 |
| 60 | 16110 | 19332 | 24165 | 32220 | 48330 | 96660 |
| 70 | 21213.3 | 25456 | 31820 | 42426.7 | 63640 | 127280 |
| 80 | 28146.7 | 33776 | 42220 | 56293.3 | 84440 | 168880 |
| 90 | 39626.7 | 41352 | 60440 | 76253.3 | 110880 | 231760 |
| 100 | 49423.3 | 59308 | 74135 | 98846.7 | 148270 | 296540 |
| | Affinity-based Delays (milliseconds) | | | | | |
| 10 | 266.667 | 320 | 400 | 533.333 | 800 | 1600 |
| 20 | 1633.33 | 1960 | 2450 | 3266.67 | 4900 | 9800 |
| 30 | 3850 | 4620 | 5775 | 7700 | 11550 | 23100 |
| 40 | 8400 | 10080 | 12600 | 16800 | 25200 | 50400 |
| 50 | 13666.7 | 16400 | 20500 | 27333.3 | 41000 | 82000 |
| 60 | 19900 | 23880 | 29850 | 39800 | 59700 | 119400 |
| 70 | 24466.7 | 25760 | 32200 | 42933.3 | 74400 | 128800 |
| 80 | 30866.7 | 35840 | 44800 | 59733.3 | 89600 | 179200 |
| 90 | 45850 | 50020 | 62275 | 84700 | 122550 | 250100 |
| 100 | 54500 | 65400 | 81750 | 109000 | 163500 | 327000 |

Table 12 shows a comparison between the greedy approach and the affinity-based approach with larger input data sizes. We consider up to 1000 user clusters. For better understanding, graphs for both the heuristic results against the number of user clusters at traffic loads of 50% and 70% are plotted in Fig. 17. Notice that affinity-based approach outperforms the greedy approach. For example, for cluster size of 100 at 40% traffic load, the total delays observed using Greedy approach are 324.5 seconds; while using Affinity–based approach the delays are 139.5 seconds. We include bars in our graphs to indicate the values of standard deviation, for better understanding. We also include Table 13 and Table 14 displaying the values of the standard deviation and margin of error (at a confidence interval of 95%) for the results presented in Table 12 to gain further insights into the performance of the heuristics.

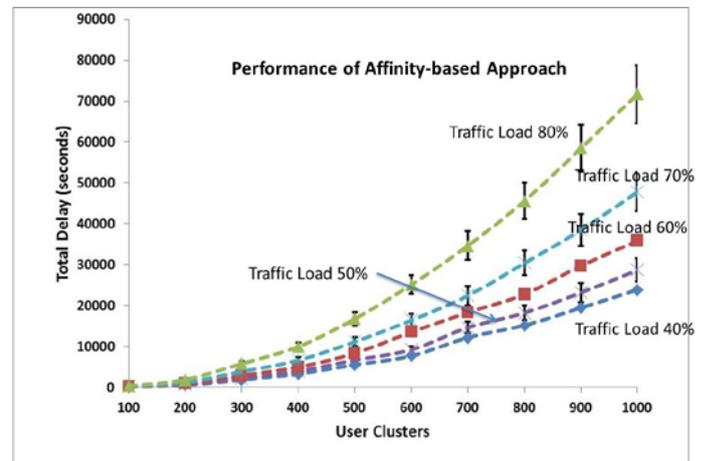

Fig. 16. Performance of Affinity-based approach

We observe better performance by the affinity-based approach even at larger topologies and higher traffic loads as well. For example, for 1000 user-clusters at 40% traffic load the total delays observed using a greedy FFD approach are approximately 50K seconds; while using Affinity–based approach, the delays are 25K seconds - an improvement of almost 50%. We present a column chart as well for the comparison between the two heuristic approaches as displayed in Fig. 18. We plot the charts for traffic loads of 60% to 80%. As discussed earlier, the improvement in the results with Affinity-based approach is clearly visible in the figure.

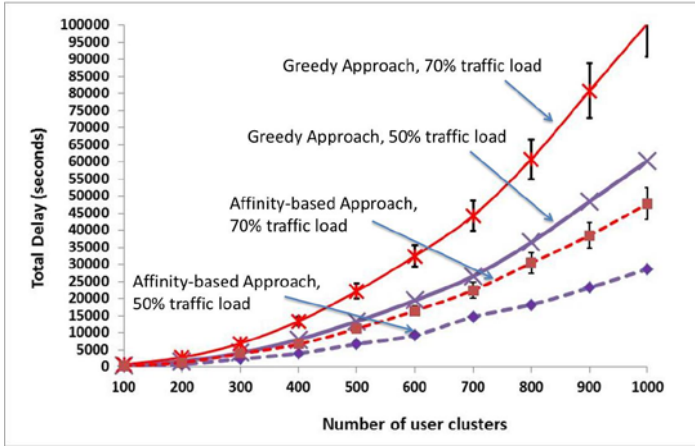

Fig. 17. FFD Greedy vs. Affinity-based heuristic.

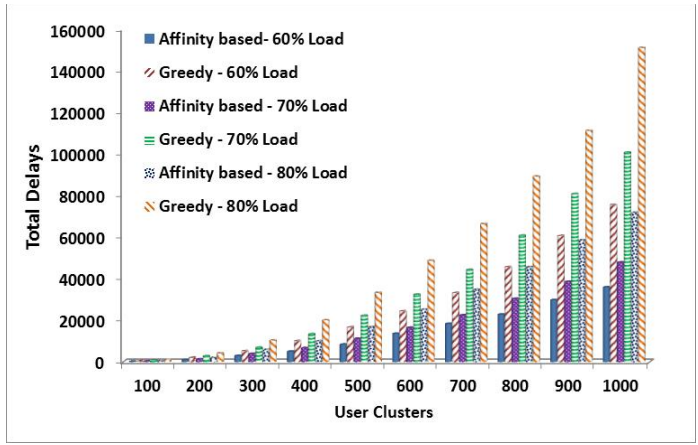

Fig. 18. FFD Greedy vs. Affinity-based heuristic – column-chart.

In Fig. 19, we plot the graphs for the total costs of the resources needed to satisfy all the given demands using both the approaches. The cost has been calculated for an hour to host the required VNFs for all the users. We assume the Amazon pricing model as shown in Table 9 to calculate the costs [50]. VNF requests are mapped to the closest matching VM from Table 9. We observe that the proposed affinity-based approach performs better than the greedy approach in terms of the total cost as well. For example, at 800 user-cluster size, the total cost to host all the required VMs for one hour using the greedy FFD approach is 65K USD while the cost using proposed ABA approach is around 40K USD. The cost difference goes on increasing with the increase in the total number of users. This may be attributed to the fact that, in the affinity-based approach, we try to accommodate the VNFs with affinity on a single VM with the closest match for the required capacities (Table 9). This reduces the required number of the resources and eventually reduces the cost.

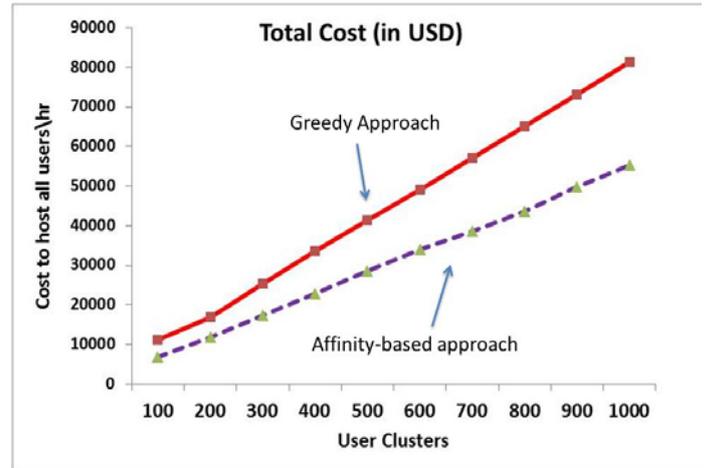

Fig. 19. FFD Greedy vs. Affinity-based heuristic – Cost comparison

TABLE 12
COMPARISON OF HEURISTIC RESULTS

| Number of User Clusters | Traffic Load (%) | | | | | |
|---|---|---|---|---|---|---|
| | 40 (%) | 50(%) | 60(%) | 70(%) | 80(%) | 90(%) |
| | FIFO Greedy Delays (seconds) | | | | | |
| 100 | 324.5 | 389.4 | 486.75 | 649 | 973.5 | 1947 |
| 200 | 1410.67 | 1692.8 | 2116 | 2821.33 | 4232 | 8464 |
| 300 | 3450 | 4140 | 5175 | 6900 | 10350 | 20700 |
| 400 | 6689.33 | 8027.2 | 10034 | 13378.7 | 20068 | 40136 |
| 500 | 11102.5 | 13323 | 16653.8 | 22205 | 33307.5 | 66615 |
| 600 | 16235 | 19482 | 24352.5 | 32470 | 48705 | 97410 |
| 700 | 22128.2 | 26553.8 | 33192.3 | 44256.3 | 66384.5 | 132769 |
| 800 | 30397.3 | 36476.8 | 45596 | 60794.7 | 91192 | 182384 |
| 900 | 40298.4 | 48397.8 | 60611 | 80804.2 | 111192 | 240920 |
| 1000 | 50316.7 | 60380 | 75475 | 100633 | 150950 | 301900 |
| | Affinity-Based Delays (seconds) | | | | | |
| 100 | 139.5 | 167.4 | 209.25 | 279 | 418.5 | 837 |
| 200 | 642 | 770.4 | 963 | 1284 | 1926 | 3852 |
| 300 | 1946.5 | 2335.8 | 2919.75 | 3893 | 5839.5 | 11679 |
| 400 | 3334.67 | 4001.6 | 5002 | 6669.33 | 10004 | 20008 |
| 500 | 5597.5 | 6717 | 8396.25 | 11195 | 16792.5 | 33585 |
| 600 | 7717 | 9260.4 | 11575.5 | 15434 | 23151 | 46302 |
| 700 | 12229 | 14674.8 | 18343.5 | 24458 | 36687 | 73374 |
| 800 | 15202.7 | 18243.2 | 22804 | 30405.3 | 45608 | 91216 |
| 900 | 19497 | 23193.3 | 29745.5 | 38501.3 | 58491 | 117982 |
| 1000 | 23908.3 | 28690 | 35862.5 | 47816.7 | 71725 | 143450 |

TABLE 13
STANDARD DEVIATION FOR RESULTS IN TABLE 12.

| Number of User Clusters | Traffic Load (%) | | | | | |
|---|---|---|---|---|---|---|
| | 40 (%) | 50(%) | 60(%) | 70(%) | 80(%) | 90(%) |
| | FIFO Greedy Delays (seconds) | | | | | |
| 100 | 31.205 | 37.046 | 47.8075 | 62.41 | 92.615 | 191.23 |
| 200 | 139.96 | 165.352 | 201.44 | 270.92 | 416.88 | 812.76 |
| 300 | 343.5 | 413.6 | 502.75 | 675 | 987.5 | 2039 |
| 400 | 637.04 | 772.448 | 957.06 | 1297.08 | 1975.12 | 3822.24 |
| 500 | 1077.23 | 1294.07 | 1594.84 | 2147.45 | 3203.68 | 6361.35 |
| 600 | 1604.15 | 1922.38 | 2327.73 | 3114.3 | 4642.45 | 9560.9 |
| 700 | 2107.53 | 2598.84 | 3157.3 | 4312.07 | 6514.61 | 12816.2 |
| 800 | 2997.76 | 3639.91 | 4501.64 | 6043.52 | 8917.28 | 17645.6 |
| 900 | 3888.85 | 4804.81 | 5953.99 | 7849.38 | 10997.3 | 22892.8 |
| 1000 | 4932.5 | 6034.2 | 7414.75 | 9862 | 14388.5 | 29283 |
| | Affinity-Based Delays (seconds) | | | | | |
| 100 | 13.555 | 16.066 | 20.8325 | 27.11 | 40.665 | 80.33 |
| 200 | 62.78 | 75.336 | 94.67 | 123.56 | 186.34 | 374.68 |
| 300 | 191.185 | 231.222 | 283.778 | 382.37 | 555.555 | 1127.11 |
| 400 | 332.12 | 386.144 | 494.18 | 651.24 | 992.36 | 1976.72 |
| 500 | 550.775 | 640.53 | 825.663 | 1064.55 | 1617.33 | 3338.65 |
| 600 | 751.53 | 901.436 | 1131.8 | 1526.06 | 2304.59 | 4575.18 |
| 700 | 1185.61 | 1410.73 | 1786.92 | 2379.22 | 3539.83 | 7171.66 |
| 800 | 1445.24 | 1798.89 | 2253.36 | 2938.48 | 4558.72 | 8961.44 |
| 900 | 1883.73 | 2240.4 | 2851.1 | 3684.12 | 5606.19 | 11748.4 |
| 1000 | 2277.75 | 2844.1 | 3430.63 | 4659.5 | 6898.25 | 13742.5 |

TABLE 14
MARGIN OF ERROR FOR RESULTS IN TABLE 12.

| Number of User Clusters | Traffic Load (%) | | | | | |
|---|---|---|---|---|---|---|
| | 40 (%) | 50(%) | 60(%) | 70(%) | 80(%) | 90(%) |
| | FIFO Greedy Delays (seconds) | | | | | |
| 100 | 1.9355 | 2.29779 | 2.96528 | 3.871 | 5.74447 | 11.8611 |
| 200 | 8.68106 | 10.256 | 12.4944 | 16.8039 | 25.8571 | 50.4117 |
| 300 | 21.3057 | 25.6537 | 31.1832 | 41.8671 | 61.25 | 126.47 |
| 400 | 39.5126 | 47.9113 | 59.3619 | 80.4518 | 122.507 | 237.076 |
| 500 | 66.8152 | 80.2651 | 98.9203 | 133.196 | 198.709 | 394.565 |
| 600 | 99.4979 | 119.236 | 144.378 | 193.165 | 287.949 | 593.018 |
| 700 | 130.721 | 161.194 | 195.833 | 267.458 | 404.07 | 794.929 |
| 800 | 185.937 | 225.767 | 279.216 | 374.851 | 553.097 | 1094.47 |
| 900 | 241.207 | 298.02 | 369.298 | 486.86 | 682.11 | 1419.93 |
| 1000 | 305.94 | 374.273 | 459.902 | 611.694 | 892.451 | 1816.29 |
| | Affinity-Based Delays (seconds) | | | | | |
| 100 | 0.84075 | 0.9965 | 1.29214 | 1.68151 | 2.52226 | 4.98249 |
| 200 | 3.89395 | 4.67274 | 5.87194 | 7.66385 | 11.5578 | 23.2396 |
| 300 | 11.8583 | 14.3416 | 17.6014 | 23.7166 | 34.4585 | 69.9094 |
| 400 | 20.5998 | 23.9507 | 30.6517 | 40.3934 | 61.5514 | 122.607 |
| 500 | 34.162 | 39.7291 | 51.212 | 66.0291 | 100.315 | 207.081 |
| 600 | 46.6139 | 55.9119 | 70.1999 | 94.6544 | 142.943 | 283.777 |
| 700 | 73.5378 | 87.5011 | 110.834 | 147.572 | 219.559 | 444.824 |
| 800 | 89.6415 | 111.577 | 139.765 | 182.26 | 282.756 | 555.836 |
| 900 | 116.839 | 138.961 | 176.84 | 228.509 | 347.726 | 728.697 |
| 1000 | 141.278 | 176.406 | 212.786 | 289.007 | 427.866 | 852.383 |

## 6. CONCLUSIONS

With cloud computing reaching the maturity, network and application service providers are looking at clouds for placing some or all of their functions in a bid to obtain flexibility while introducing new services. This has led to a recent spurt in interest in service function chaining and network function virtualization. In this work, we have presented an analytical study of these two concepts with current research directions, especially the problem of placing service function chains over the network function virtualized platform in a multi-cloud scenario. The focus of the work is on reducing the total delays to the end users and total cost of deployment for service providers in inter-cloud environments. To achieve this, we aim to reduce the inter-cloud traffic between virtual function instances, flowing through the service chains. We have considered cost constraints as well as other SLA constraints while formulating the model.

We formulate an optimization model with applicable constraints. The problem has been solved using an Integer Liner Programming (ILP) methodology. It has been observed that because of computational complexity, the ILP model has limited applicability, especially to the cases with a small number of user nodes. To overcome this limitation, we propose a novel Affinity-based approach (ABA). We have considered different user-levels with different user delay tolerances. We also satisfy QoS as well as placement related SLAs. In addition, the traffic-affinity between the VNFs has been taken into consideration for their placement in the clouds. We provide a performance comparison between the proposed ABA heuristic and simple greedy approach (SBA) using first-fit decreasing (FFD) method, which has already been widely studied in the literature for the VM placement problem. We present results for both the heuristics and observe that the quality of the solution is much improved using Affinity-based approach with only a marginal increase in execution time as compared to the FFD greedy approach. We believe that the proposed work may be extended to accommodate more complex SLAs and QoS constraints in the future.

### ACKNOWLEDGEMENT

This work has been supported under the grant ID NPRP 6 - 901 - 2 - 370 for the project entitled "Middleware Architecture for Cloud Based Services Using Software Defined Networking (SDN)", which is funded by the Qatar National Research Fund (QNRF) and a grant from Huawei. H. Anthony Chan was supported by Huawei. The statements made herein are solely the responsibility of the authors.